# Enhancing Data Storage Reliability and Error Correction in Multilevel NOR and NAND Flash Memories through Optimal Design of BCH Codes

Saeideh Nabipour
Department Computer and Electrical Engineering
University of Mohaghegh Ardabili
Ardabil, IRAN
Corresponding Author's Email:
Saeideh.nabipour@gmail.com

Javad Javidan
Department Computer and Electrical Engineering
University of Mohaghegh Ardabili
Ardabil, IRAN
javidan@uma.ac.ir

**Abstract**: The size reduction of transistors in the latest flash memory generation has resulted in programming and data erasure issues within these designs. Consequently, ensuring reliable data storage has become a significant challenge for these memory structures. To tackle this challenge, error-correcting codes like BCH (Bose-Chaudhuri-Hocquenghem) codes are employed in the controllers of these memories. When decoding BCH codes, two crucial factors are the delay in error correction and the hardware requirements of each sub-block. This article proposes an effective solution to enhance error correction speed and optimize the decoder circuit's efficiency. It suggests implementing a parallel architecture for the BCH decoder's sub-blocks and utilizing pipeline techniques. Moreover, to reduce the hardware requirements of the BCH decoder block, an algorithm based on XOR sharing is introduced to eliminate redundant gates in the search Chien block. The proposed decoder is simulated using the VHDL hardware description language and subsequently synthesized with Xilinx ISE software. Simulation results indicate that the proposed algorithm not only significantly reduces error correction time but also achieves a noticeable reduction in the hardware overhead of the BCH decoder block compared to similar methods.

## 1. Introduction

Flash memories emerged as the initial formidable rival to hard disks. These semiconductor memories function similarly to non-volatile hard drives; however, their access speed is significantly faster, ranging from 100 to 1000 times quicker than traditional hard drives. In terms of their physical dimensions, they are significantly smaller. When it comes to energy consumption, they are more cost-effective and less vulnerable to physical damage. Flash memories have the capability to store data on a silicon chip, eliminating the need for continuous power supply. Essentially, flash memories can be categorized as a form of EEPROM (Electrically Erasable Programmable Read-Only Memory) memory, which allows for electrical programming and erasing of data. The producers of flash memory devices have made significant investments to enhance technology and meet the growing demand in the market. Flash memory is extensively utilized in various devices such as mobile phones, digital cameras, MP3 players, and solid-state drives (SSDs), contributing to its widespread popularity [1].

Flash memories have been developed using various techniques and technologies. The initial version, called Flash NOR, had individual cells resembling standard NOR gates. Later, Flash NAND memories were introduced, offering significantly higher density and storage capacity. However, there are differences in how data is accessed and written between the two types. In NAND memories, data is read and written in blocks, and specific wiring is required for accessing any particular location in the memory space. Additionally, when data is erased in NAND memories, it happens at the block level, not at specific points within the memory. Flash memory is manufactured using two distinct cell structures: single-level cell (SLC) and multi-level cell (MLC). MLC-based memories have the capacity to store 2 bits of information in each memory cell, while SLC chips can only store 1 bit of data in each memory cell.

Flash memories that are based on a cell structure with multiple levels can store n bits of information by creating $2^n$ states. Each state corresponds to a specific voltage range, representing different voltage levels. Currently, the development of flash memory technology focuses on increasing storage capacity and reducing production costs, leading to a trend of making them smaller. The use of very small MOSFETs in flash cells is highly desirable due to their advantages. However, reducing the size of flash memory cells also decreases the distances between adjacent levels, which can cause voltage fluctuations when reading and writing data. This issue becomes more significant when utilizing multi-level cells, requiring precise and reliable data storage techniques.

Transistors with small sizes do not have the capability to handle high voltages effectively when it comes to writing and erasing data in flash memory. The presence of a high charge in the delicate and brittle oxide insulation between the floating gate and the underlying layer, as well as between the floating gate and the source, disrupts the state of the floating gate electrons, causing issues with accurate programming and data erasure, leading to errors in flash memory. Meanwhile, the technology for reducing the physical structure of MOSFET transistors and high-speed flash memories is advancing, but it has also made the production of these memory types quite challenging, limited to only a few countries worldwide [2, 1].

Bit errors in flash memories can be divided into two categories: soft errors and hard errors. Hard errors occur when the oxide in the flash memory collapses due to limitations in writing or erasing, resulting in permanent bit failures. On the other hand, soft errors stem from various mechanisms, such as disruptions in the read and write processes and data retention, but they can be resolved in the subsequent write/erase cycle. In the field of multi-level cell flash memory technology, which faces significant challenges in data storage reliability, error control methods are utilized. Generally, to tackle this type of challenge in error control for these memories, an ECC block is integrated. Its primary purpose is to overcome high error rates and enhance storage reliability by employing signal processing techniques like error correction codes (ECC) capable of rectifying multiple bit errors.

Error correction codes enable the recipient to verify transmitted data without needing to request a resend of the information. Initially, they add extra redundancy to the sent message through encoding, which occurs at the sender's side. Subsequently, any errors that occur during transmission can be rectified through decoding, which involves extracting the added redundancy at the receiver's end. The result of the encoding process is a code word containing both the original message and the added redundancy known as a parity bit. In smaller flash memories, a page is the smallest unit that can be read or written coherently and is independently encoded/decoded using an ECC (Error Correction Code) block. During the writing phase, the ECC block encodes the data from the I/O interface and writes it to the designated page. During reading, the ECC block decodes the data of each page from the memory and sends it to the I/O interface.

BCH codes are commonly used in flash memories to correct errors, and they are known for their effectiveness in handling complex error correction tasks, particularly for multiple errors. These codes were initially discovered by Hocquenghem in 1959 and later independently proven by Bose and Chaudhuri in 1960 [3]. The error correction process using a BCH decoder involves four steps: calculating the syndrome, determining error locator polynomials, employing the search algorithm of Chein to locate the errors, and finally correcting the errors. Over time, various efforts have been made to optimize BCH decoders. Additionally, different algorithms have been proposed to reduce decoding latency and minimize the hardware required for the decoder block. The implementation of a BCH decoder involves a trade-off between decoding time and hardware size, and most approaches aim to strike a balance between the two, considering the necessary time for error correction and the size of the decoder circuit.

In reference [4], a specific algorithm is introduced to decrease the hardware complexity resulting from the multiplication process based on prime field polynomials in finite field multiplier blocks. This algorithm employs the parallel search Chien operation alongside shift operations. Its main goal is to not only reduce the number of multipliers but also minimize the number of adders. Another design named MPCN-based parallel architecture is presented in reference [5], which utilizes combination techniques for substituting finite field multipliers. The proposed algorithm formulates the search Chien equation by employing mini polynomial combinations. One notable feature of this algorithm is its ability to integrate the syndrome generation and search Chien blocks with the least possible hardware complexity. In reference [6], the aim is to reduce the hardware size of the syndrome generation block by utilizing the "on-demand" approach. According to the proposed method, only odd syndromes are generated in the syndrome block. However, if even syndromes are needed in the error locator polynomial generation block, they can be calculated using the property $S_{2i} = S_i^2$.

In reference [7], a technique known as re-encoding is employed to minimize the decoding process delay. This approach involves dividing the syndrome calculation into two phases: error detection and identification of error correction components. The suggested method assigns the initial phase to the encoder block, which remains inactive during the reading of idle data. Consequently, it can detect errors without requiring any additional hardware, solely relying on LFSR (Linear Feedback Shift Register). Furthermore, an algorithm called oBM is recommended for generating error locator polynomials, enabling the execution of concurrent instructions. In references [8, 9], alternative algorithms such as RiBM, SiBM, and TiBM are proposed as well, aiming to generate error locator polynomials with reduced hardware complexity and faster processing in clock cycles.

Two crucial aspects that receive attention in the decoding process of BCH codes, besides error correction capability, are the time delay and hardware volume of each sub-block. The objective of this article is to tackle these challenges and present optimal methods and algorithms that effectively decrease the decoding time while avoiding an increase in hardware complexity, thus striking a balance between these two factors. Considering the widespread utilization of BCH error correction codes in flash memories to enhance data storage reliability, it is deemed important to propose optimization approaches that prioritize both time and hardware complexity parameters simultaneously during the decoding process.

This article presents a technique to decrease decoding latency in the process of decoding BCH sub-blocks. It utilizes parallelism and introduces an XOR-repeated sharing algorithm to reduce the hardware requirements of the BCH decoder block. To showcase the superiority of this approach, different sets of data with varying error counts are applied to the BCH decoder block, and the outcomes are compared to similar methods. The results validate the anticipated improvements achieved through the implemented architectures and techniques.

The article is structured as follows: In Section (2), it provides a description of BCH error correction codes and examines the performance of the encoder and decoder blocks in controlling flash memory errors. Section (3) introduces a parallel architecture for the BCH decoder block, aiming to reduce decoding process delay. Subsequently, Section (4) thoroughly explains the algorithm for XOR sharing in the BCH decoder block, aiming to minimize hardware complexity. Section (5) presents a technique known as pipeline to enhance the efficiency of the BCH decoder circuit. Section (6) focuses on explaining and analyzing the results obtained from simulations. Finally, in Section (7), the article concludes and summarizes the key findings.

## 2. Explanation of BCH error correction codes

So far, there have been different types of error correction codes suggested for correcting errors in flash memories. These codes include Hamming codes, Solomon-Reed codes, BCH codes, Goppa codes, Golay codes, and more. While the effectiveness of codes like Goppa codes in error correction has been demonstrated [2], manufacturers of flash memories, considering the wide range of these memories and the need for highly accurate error correction mechanisms with maximum efficiency and minimal hardware complexity, have embraced the adoption of codes that possess high error correction capabilities and optimized hardware implementation.

Older flash memory chips utilized Hamming codes based on SLC (Single-Level Cell) for error correction. These codes, despite their simplicity, could only rectify a single random error, rendering them ineffective for practical use in flash memory. With the increased error rates in newer generations of SLC and MLC (Multi-Level Cell) flash memory, designers now employ more robust and intricate codes like Solomon-Reed and BCH (Bose-Chaudhuri-Hocquenghem) codes. These codes belong to the family of linear block codes and are a subset of cyclic codes. While RS (Reed-Solomon) codes perform error correction on binary symbols and demonstrate subpar performance in the presence of current noise, BCH codes excel at error correction on byte-level symbols and offer superior resilience against random noise compared to RS codes.

The analysis conducted on flash memories reveals that the errors observed in each memory page are random in nature. This is why BCH codes are widely employed in flash memories [10]. BCH codes are constructed based on the extension of the Galois field, denoted as GF ($2^m$), and they utilize finite field calculations with a finite field size of m for encoding and decoding. To put it simply, GF ($2^m$) encompasses all possible combinations of m bits, represented as powers of α or the primary element of the Galois field GF(m). The Galois field GF($2^m$) is formed by utilizing irreducible polynomials of degree m with α as its root. For any positive integer m (m ≥ 3), there exists a binary BCH code with the parameters indicated in Table 1 [3].

Table: [1] Parameters of BCH Error Correction Code [3]

| | |
|---|---|
| The size of the code template | $n = 2^m - 1$ |
| Redundancy Bit | $n - k \leq mt$ |
| Minimum Distance | $d_{min} \geq 2t + 1$ |
| The capacity to rectify mistakes | $t = \left\lceil \frac{m-k}{2} \right\rceil$ |

Table 1 illustrates the parameters of the BCH code, which include various values: n represents the total number of bits, including the additional parity bits; k denotes the length of the message; t signifies the ability to correct m errors; m refers to the size of the finite field GF ($2^m$); and $d_{min}$ represents the minimum distance between two code words. The message length is primarily used for flash memories and is typically a power of 2. Since the size of a page or sector in a flash memory may not align with the parameter k of a BCH code, shortened BCH codes are employed for error correction. Shortened codes BCH (n-l, k-l, t) are obtained by removing l zero bits, and they possess the same error correction capability as the original BCH code [11].

The Flash NAND memory bank is composed of multiple blocks, with each block containing 64 pages for SLC memory and 128 pages for MLC memory. Each memory block holds 2048 bytes of data, along with an additional 64 bytes of information. Within a Flash NAND memory block, each page is divided into four sectors of 512 bytes, and each sector includes 16 extra bytes. Therefore, considering the condition $2^{12} \leq (512 + 16) \times 8 \leq 2^{13} = 8192$, these memory types operate using GF ($2^{13}$) field. To handle potential errors in each memory page of the 13 GF memory, a shortened code (4096, 420) BCH is employed for the simultaneous reading of 512 bytes (equivalent to 4096 bits) from a memory array, under GF ($2^{13}$) field. As for Flash NOR memories, which read a data queue of 256 bits (16 groups of 16 bits) simultaneously, a shortened code (256, 274) BCH under GF ($2^9$) is utilized for error correction in each memory page. [6].

BCH codes have the ability to correct *t* or fewer errors in a block of length n=$2^m$-1 bits. The number of generator polynomials for these codes, which have *m* roots belonging to the Galois field GF($2^m$), can be determined. The element α is identified as the primitive element of GF($2^m$). The generator polynomials for the BCH error-correcting code, with a length of g(x) and the minimum degree under the field GF (2), can be calculated. The roots α, $α^2$, ..., $α^{2t}$ correspond to the powers of α where ($g(α^i)_{1≤i≤2t}$ = 0) holds true. By considering the minimum degree polynomials under the field GF(2) as $φ_1(X)$, $φ_2(X)$, ..., $φ_{2t}(X)$, the element $α^i$ represents the smallest non-zero multiple of (g (x)) [3].

$$g(X) = LCM\{\phi_1(X), \phi_2(X), \ldots, \phi_3(X)\} \tag{1}$$

The ECC block in the flash memory controller is comprised of two primary modules: an encoder module and a decoder module. When *k* bits of user data are being written to flash memory, an encoder circuit generates additional parity bits. These parity bits are added to the *k* bits of data, resulting in the creation of an *n*-bit codeword. Subsequently, the entire codeword is stored and written to a page in the memory bank. While reading, the decoder circuit scans a code word for errors and, depending on its error correction capability, rectifies any inaccuracies within the code word. Consequently, the accurate code word is recovered.

## 2.1. Block BCH Encoder

The act of encoding takes place within multiple generator sentences. The encoding process in the ECC flash memory controller block involves converting a k-bit data into a codeword of n bits, represented as ($c_1, c_2, \ldots, c_n$), ($c_i \in$ GF(2)).

1- Move v(x) towards the left by shifting it n - k bits, resulting in the creation of n - k bits with a value of zero on the right side. One way to achieve this outcome is by taking the product of v(x) and $X^{n-k}$, and then presenting the value as $X^{n-k}$v(X).
2- The calculation of the remaining polynomials resulting from the previous stage, in relation to the generating polynomials (degree of redundancy).

$$r(X) = X^{n-k}v(X) \bmod g(X) \tag{2}$$

3- To produce a code word called C(X) consisting of n bits, $X^{n-k}v(X)$ is positioned in an available space on the right side of *r(X)* within *(n-k)*.

$$C(X) = V(x)x^{n-k} + Re\ m(V(X)x^{n-k})g(x) \tag{3}$$

4- The primary component of an encoder block is the Linear Feedback Shift Register (LFSR), which is depicted in Figure 1 [14]. The increase in the number of bits is achieved by utilizing a shift register of length (n - k), where (n - k) represents the degree of the generating polynomial. This expansion is accomplished by connecting the relevant feedback associated with the coefficients of the generating polynomial.

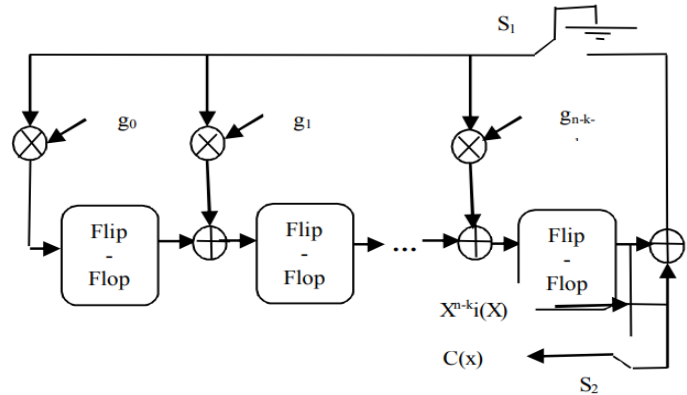

Figure 1 - Block Diagram of BCH Encoder

## 2.2. BCH decoder block

Figure 2 depicts the four stages involved in the BCH code, namely syndrome calculation, error locator polynomial calculation, the Chien search algorithm for locating the errors, and their subsequent correction. Repeatedly performing the write/erase cycle on flash memory, considering its inherent limitations in data storage cycles, has the potential to harm the stored information, leading to read errors. If the code word that is kept in memory follows the pattern of $c(X) = c_0 + c_1 X + c_2 X^2 + \cdots + c_{n-1} X^{n-1}$, and the transfer is carried out based on equation (4), it leads to the reception of the code word r(x).

$$r(X) = r_0 + r_1 X + r_2 X^2 + \cdots + r_{n-1} X^{n-1} \tag{4}$$

Once the error vector e is combined with the encoded code vector c, the resulting received vector r can be expressed using equation (5). Additionally, there will be several error polynomials of degree $e(x) = e_0 + e_{1x} + \cdots + e_{n-1} x^{n-1}$ included in the representation of r.

$$r(X) = c(X) + e(X) \tag{5}$$

Afterward, the task of deciphering and rectifying errors is carried out through the utilization of the BCH decoder block.

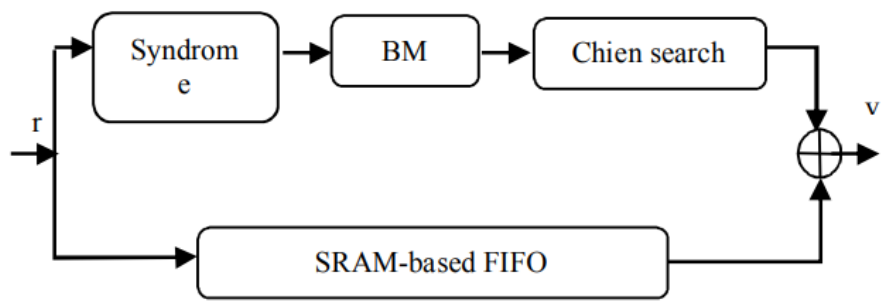

Figure: 2 Block Diagram Decoder for BCH

## Stage One: Calculation of Syndrome

The initial phase of decoding a BCH code involves determining errors by calculating the syndrome using the received vector (X). In the coding part of BCH codes, it is evident that each code word is generated by multiplying $X^{n-k}v(X)$ and taking its remainder with respect to g(x). Consequently, the resulting polynomial from the encoder block does not contain any remainders in relation to $X^{n-k}v(X)$. This is because, during the encoding process, the remainder $X^{n-k}v(X)$ is incorporated into it. Therefore, when we divide the resulting sentences of x(g), we can determine if an error exists. If the received data is divisible by a few generating sentences, it indicates that no error has occurred. Consequently, we can infer that the received codeword is invalid and contains an error. To calculate 2t, which represents the syndrome component, we utilize equation (6).

$$S(X) = (S_1, S_2, \ldots, S_{2t})$$

$$S_i = r(\alpha^i) = e(\alpha^i) = \sum_{j=0}^{n-1} r\alpha^{ij} \tag{6}$$

$$1 \leq i \leq 2t$$

### Stage Two: Finding Error-Detector Polynomial

In the decoding process of the BCH decoder block, the second step involves the identification of error-locator polynomials σ (X), as stated by equation (7). This step of the decoding process involves the utilization of different algorithms, which can be classified into three main groups: the Paterson algorithm, the Euclidean algorithm, and the Berlekamp-Massey (BM) algorithm.

$$\sigma(X) = \sigma_0 + \sigma_1 X + \sigma_2 x^2 + \cdots + \sigma_v X^v \tag{7}$$

Although the Paterson algorithm is conceptually straightforward, it becomes more complex due to a growing number of correction candidates (t) within a codeword. The algorithm's computational complexity increases as the value of t rises, resulting in higher intricacy in its calculations. Hence, more advanced algorithms like BM and Euclid are preferred for finding error-correcting polynomials. In this article, we utilize the BM algorithm to identify error-correcting polynomials.

### Stage Three: Utilizing the Chien search algorithm to locate the position of the error

After successfully identifying the error locator polynomial X(σ), the next task is to locate where the error occurred. By inverting and expressing the values of X(σ) using α powers, we can determine the positions of the errors. To simplify this process, the Chien search algorithm can be utilized. The roots of the error-locator polynomial X(σ) are derived from elements within the Galois field specific to the BCH code being used. To find these roots, it is necessary to evaluate the value of X(σ) for each field element. Any values that yield a result of zero are considered as one of the roots of X(σ), referred to as (σ ($\alpha^i$)=0).

## Stage Four: Error Correction

In the last step, to obtain the correct code word, we combine the word with errors X(c) with the output of the search Chien block. This is done by adding the bits of the code word stored in the corresponding positions of the FIFO (First in First Out) buffer. Consequently, the error correction block consists of a single finite field adder (FFA), which is structured based on XOR gates.

$$V(X) = C(X) + e(X) \tag{8}$$

## 3. Parallel Decoders Architecture for BCH

The BCH decoder block encounters certain constraints within its serial architecture, resulting in longer computation times, particularly in MLC Flash technology where data is written in blocks, usually 512 bytes in size. This delay becomes quite noticeable. To address this issue, an efficient approach based on parallel architecture has been implemented for each of the sub-blocks responsible for generating syndromes and conducting search Chien. This implementation reduces the time delay significantly and speeds up the overall computation process [16]. Moving forward, we will discuss the introduction of parallel architecture for each of the sub-blocks involved in syndrome generation and Chien search.

The architecture for parallel syndrome generation involves a block diagram that utilizes a finite-field multiplier, a register with p bits, and a finite-field adder. This design allows for the simultaneous computation of syndromes on all the input bits, which are received in a single clock pulse. By employing the parallel-p architecture, the delay is reduced from n pulses to $\lceil n/p \rceil$ pulses.

One aspect to consider is the connection between the even and odd syndromes in $S_{2i} = S_i^2$ binary BCH codes [3]. By utilizing the odd syndrome components and a squaring circuit composed of XOR gates, it becomes straightforward to calculate the removed even syndromes and their components directly. This approach effectively reduces the hardware complexity of the parallel architecture in the syndrome calculation block [17].

Parallel Architecture Block Search Chien utilizes a parallel approach to find the roots of error locator polynomials. In the serial block search Chien architecture, the evaluation of n elements from the Galois field GF($2^m$) takes n clock cycles. However, in the parallel block search Chien architecture, the evaluation of $[\sigma(\alpha^i), \sigma(\alpha^{2i}), ..., \sigma(\alpha^{pi})]$ p elements from the Galois field GF($2^m$) occurs simultaneously in a single clock cycle, resulting in a significant reduction in computation delay from n clock cycles to $\lceil n/p \rceil$ clock cycles. In parallel architecture, the Chien search block is utilized in a combination of components including a p-finite field multiplier with dimensions (p × t), a finite field adder with m bits and t inputs, a register with p rows and m bits for finite field subtraction, and a multiplier with p rows and m bits for finite field multiplication. It's worth mentioning that m and t represent the expansion degree of the Galois field GF($2^m$) and the error-correction capability of BCH code respectively [20-18].

## 4. The XOR-sharing algorithm to reduce the size and hardware complexity of the B decoder block

It is important to note that employing parallel techniques necessitates more intricate and larger hardware components. Consequently, in the design phase of the BCH decoder block, our objective is to minimize the hardware requirements and overall size. The Chien search block, out of the three blocks in the BCH decoder, has the largest hardware size and computational complexity. It accounts for roughly 65% of the overall volume of the BCH decoder block in error

(9)

correction systems [4, 5]. The search Chien algorithm assesses the error location using an $\sigma(\alpha^i)_{0 \leq i \leq n-1} = 0$ evaluation. Typically, an error-detector polynomial can be expressed as equation (9).

$$\sigma(X) = \sigma_0 + \sigma_1 X + \sigma_2 X^2 + \cdots + \sigma_t X^t$$

$$\sigma(\alpha^i) = \sigma_0 + \sigma_1 \alpha^i + \sigma_2 \alpha^{2i} + \cdots + \alpha_t \alpha^{ti}$$

Equation (9) indicates that in order to compute its value, we need to multiply it by several constant finite field multipliers (CFFM) and the value of 12. Consequently, the most utilized component in the internal structure of the search block Chien is the CFFM, resulting in increased hardware complexity when implementing the circuit in parallel. To mitigate this complexity, we can consider the product of m-bit multiplicative constants (where $1 \leq j \leq t$) in the polynomial field as a constant $\alpha_{1 \leq i \leq p}^{ij}$, as stated in equation (10). (10)

$$P = \alpha^{ij}\sigma_j = \alpha^{ij}(\alpha_{j,0} + \alpha_{j,1}\alpha + \cdots + \alpha_{j,m-1}\alpha^{m-1})$$

For the purpose of performing multiplication, instead of relying on a finite field multiplier with two stages: 1) multiplying the coefficients of the multiplicand by the coefficients of the multiplier, and 2) calculating the remainder obtained from dividing the product by the primitive polynomial of the Galois field, one can opt to utilize the Mastrovito multiplier, which involves matrix multiplication. The Mastrovito multiplier effectively combines the two aforementioned stages in Galois field multipliers through a form of parallel multiplication, as described by equation (11) [21].

$$P = \alpha^i \sigma_j = \alpha^i(\sigma_{j,0} + \sigma_{j,1} + \cdots + \sigma_{j,m-1})$$

$$= \begin{bmatrix} \alpha_0^i & \alpha_0^{i+1} & \cdots & \alpha_0^{i+m-1} \\ \alpha_1^i & \alpha_1^{i+1} & \cdots & \alpha_1^{i+m-1} \\ \vdots & \vdots & \ddots & \vdots \\ \alpha_{m-1}^i & \alpha_{m-1}^{i+1} & \cdots & \alpha_{m-1}^{i+m-1} \end{bmatrix} \begin{bmatrix} \sigma_{j,0} \\ \sigma_{j,1} \\ \vdots \\ \sigma_{j,m-1} \end{bmatrix} \quad (11)$$

$$= [a_{coeff}]_{m \times m} * [\sigma_j]_{m \times 1}$$

$$= [P_{j,0} \quad P_{j,1} \quad P_{j,2} \quad \cdots \quad P_{j,m}]^T$$

Due to the fact that the $\alpha_l^{i+j}$ $(0 \leq l, j \leq (m-1))$ elements are in binary form and addition in the Galois field is defined as bitwise XOR, the computational complexity of finite field multipliers can be reduced by employing XOR gates. To simplify and decrease the hardware required for this multiplication operation even further, the XOR sharing multiplier is utilized.

This multiplier operates by identifying shared repetitive subsequences, aiming to eliminate unnecessary calculations and decrease the number of XOR operations performed. The XOR-sharing multiplier can detect redundant XORs that occur multiple times. If it becomes feasible to precompute the pairs of elements that are repeatedly multiplied in the matrix, it leads to the elimination of redundant calculations and a reduction in hardware requirements. By going through four stages, the XOR-sharing multiplier effectively removes superfluous XOR gates [22, 23].

- Stage 1: Checks every i and j column of the $a_{\text{coffe}}$ matrix and chooses the columns where the elements have more than one occurrence of a match.
- Stage 2: In the second stage, the columns that exhibit the most matches between their elements and the columns from the first stage are selected as the match-best columns.
- Stage 3: The elimination of repetitive computations resulting from the XOR gate will be achieved by adding a column to the right side of matrix $a_{\text{coffe}}$.
- Stage 4: In this stage, the process of stages 1 to 3 is iterated across all columns of the $a_{\text{coffe}}$ matrix until the occurrence of matches between the elements of column 1 and column 2 is limited to a maximum of one occurrence.

Once the four optimization stages mentioned earlier are finished, the XOR gates that occur repeatedly are chosen, computed in advance, and shared among various sentence multiplication operations. This leads to a reduction in the number of XOR gates and the hardware complexity of the finite field multiplier. Moreover, the XOR sharing multiplier enables the application of not only individual finite field multipliers within the search Chien block but also a group of multipliers with the same ($\sigma_j$) multipliers. This effectively decreases the quantity of XOR gates to a significant extent.

We are examining the execution of this multiplier for a set of multipliers in our study. Imagine the Chien search block considering the four-field finite field multiplier operating under the Galois field, specifically denoted as GF($2^9$) and expressed as $\sigma_5\alpha^5$, $\sigma_5\alpha^{10}$, $\sigma_5\alpha^{15}$, $\sigma_5\alpha^{20}$. Since each of the four multipliers has a common product of one ($\sigma_5 = B$), it is possible to apply the XOR-sharing multiplier to any of them.

To demonstrate, we can show that the result of multiplying four constant field multipliers, represented as P$_1$, P$_2$, P$_3$, and P$_4$ respectively according to equation (11), yields the following outcomes.

$$P_1 = \alpha^5 B = \begin{bmatrix} b_5 \\ b_6 \\ b_7 \\ b_8 \\ b_0 + b_5 \\ b_1 + b_6 \\ b_2 + b_7 \\ b_3 + b_8 \\ b_4 \end{bmatrix} \quad (12)$$

$$P_2 = \alpha^{10} B = \begin{bmatrix} b_0 + b_5 \\ b_1 + b_6 \\ b_2 + b_7 \\ b_2 + b_3 + b_8 \\ b_0 + b_4 + b_5 \\ b_1 + b_5 + b_6 \\ b_2 + b_6 \\ b_3 + b_7 + b_8 \\ b_4 + b_7 + b_8 \end{bmatrix} \tag{13}$$

$$P_3 = \alpha^{15} B = \begin{bmatrix} b_0 + b_4 + b_5 \\ b_1 + b_5 + b_6 \\ b_2 + b_6 + b_7 \\ b_2 + b_3 + b_7 + b_8 \\ b_0 + b_5 + b_8 \\ b_0 + b_1 + b_6 \\ b_1 + b_7 \\ b_2 + b_3 + b_8 \\ b_2 + b_3 + b_4 \end{bmatrix} \tag{14}$$

$$P_4 = \alpha^{20} B = \begin{bmatrix} b_0 + b_5 + b_8 \\ b_0 + b_1 + b_6 \\ b_1 + b_2 + b_7 \\ b_2 + b_3 + b_8 \\ b_0 + b_3 + b_4 + b_5 + b_8 \\ b_1 + b_4 + b_5 + b_6 \\ b_2 + b_5 + b_6 + b_7 \\ b_3 + b_6 + b_7 + b_8 \\ b_4 + b_7 + b_8 \end{bmatrix} \tag{15}$$

Considering the multiplication equation matrix provided, we can determine that each of the four multipliers requires a specific number of XOR gates: 4, 14, 18, and 23, respectively. By applying the XOR sharing multiplier to each multiplier and identifying any repetitive expressions, we obtain three distinct groups of XOR equations as follows:

$$P_2 : c_1 = b_7 + b_8; c_2 = b_0 + b_5; c_3 = b_1 + b_6$$
$$P_3 : c_1 = b_2 + b_3; c_2 = b_4 + b_5; c_3 = b_1 + b_6 \qquad (16)$$
$$P_4 : c_1 = b_7 + b_8; c_2 = b_0 + b_5; c_3 = b_1 + b_6$$
$$c_4 = b_2 + b_7$$

Therefore, by utilizing the shared redundant XORs in each multiplier, we can decrease the number of XOR gates needed for $P_2$ from 14 to 7, for $P_3$ from 18 to 11, and for $P_4$ from 23 to 14. This reduction is achieved by identifying and utilizing common XORs among the coefficient matrices of $P_1$ to $P_4$, resulting in a set of 15 equations:

$$
\begin{aligned}
&c_1 = b_0 + b_5; \quad c_2 = b_1 + b_6; \quad c_3 = b_2 + b_7 \\
&c_4 = b_3 + b_8; \quad c_5 = b_2 + b_6; \quad c_6 = b_7 + b_8 \\
&c_7 = b_2 + b_3; \quad c_8 = c_1 + b_4; \quad c_9 = c_2 + b_5 \\
&c_{10} = b_4 + c_6; \quad c_{11} = c_6 + b_3; \quad c_{12} = c_1 + b_8 \\
&c_{13} = b_0 + c_2; \quad c_{14} = c_2 + b_4; \quad c_{15} = b_7 + c_5
\end{aligned}
\qquad (17)
$$

Therefore, by utilizing the 15 equations obtained earlier, we can rephrase the coefficient matrix from $P_1$ to $P_4$ in the following manner:

$$P_1 = \alpha^5 B = \begin{bmatrix} b_5 \\ b_6 \\ b_7 \\ b_8 \\ c_1 \\ c_2 \end{bmatrix} \quad ; \quad P_2 = \alpha^{10} B = \begin{bmatrix} c_1 \\ c_2 \\ c_3 \\ c_{11} \\ c_8 \\ c_9 \end{bmatrix}$$

$$P_3 = \alpha^{15} B = \begin{bmatrix} c_8 \\ c_9 \\ c_{15} \\ c_6 + c_7 \\ c_{11} \\ c_{13} \\ b_1 + b_7 \\ c_{14} \\ c_4 + c_7 \end{bmatrix} \quad ; P_4 = \alpha^{20} B = \begin{bmatrix} c_8 \\ c_9 \\ c_{15} \\ c_6 + c_7 \\ c_{11} \\ c_{13} \\ b_1 + b_7 \\ c_{14} \\ c_4 + c_7 \end{bmatrix}$$

In general, it can be inferred that both multiplier types have the ability to noticeably reduce the hardware complexity of the Chien search block by utilizing repeated XORs, whether within individual multipliers or across a group of multipliers. As a result, this reduction in hardware complexity can lead to a decrease in the hardware requirements of the BCH decoder block.

## 5. Pipeline Technique in BCH Decoder Circuits to improving the Performance

In the usual scenario, the BCH decoder relies on a parallel architecture to process a code word. It receives the entire code word in $n/p_s$ clock cycles, where $p_s$ represents the parallelization factor or the number of elements processed simultaneously in the syndrome calculation circuit. During this time, it also calculates the syndrome values. Next, it executes the BM multiplier on the syndromes in 2t clock cycles, obtaining the coefficients of the error locator polynomials. Finally, using the search Chien multiplier, it determines the possibility of error occurrence in $n/p_c$ clock cycles, where pc represents the parallelization factor or the number of elements processed simultaneously in the search Chien block. The corrected code word is then outputted in the same order as it was received. Consequently, the delay of the error correction process in the BCH decoder block is equal to $n/p_s/2t+n/p_c$ clock cycles. It's important to note that after receiving each code word, the circuit needs to perform calculations for $2t+n/p_c$ clock cycles (2t cycles for BM and $n/p_c$ cycles for the search Chien multiplier). During this time, no new input should be received until the previous code word has completely exited the decoder circuit. Essentially, the circuit's timing performance will resemble Figure 3.

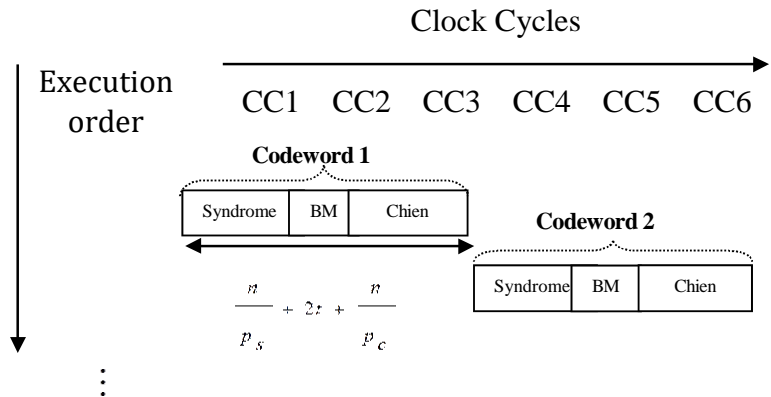

Figure 3: Non-pipeline Architecture of BCH Decoder circuits

In order to enhance the functioning of the circuit, we can utilize a pipelining technique known as "Overlapping," where multiple instructions are executed concurrently [24]. Essentially, the pipeline architecture is composed of several stages, each assigned with specific tasks that can be carried out simultaneously. In simpler terms, when processing a code word during operation, we desire the circuit's input to be capable of receiving a new code word. It is important to note that acquiring the input necessitates $n/p_s$ clock pulses, while processing it requires $2t+n/p_c$ clock pulses. Therefore, it is not possible to execute the processing of an input code word concurrently with the reception of a new code word. To put it differently, the hardware implementation cannot receive two code words in the syndrome block and perform calculations on them within a single cycle. Consequently, we need to pause the input for a duration of 2t clock pulses.

The provided paragraph discusses the architecture of a pipeline for three-stage decoders in a BCH circuit, as shown in Figure 4. This architecture enables simultaneous decoding and error correction processes on three code words. According to Figure 4, the syndrome circuit operates concurrently with the BM circuit for the first and second code words, while the syndrome circuit for the third code word operates simultaneously with the BM circuit for the second code word and the search Chien circuit for the first code word. Both architectures depicted in Figures 3 and 4 share identical hardware components. In the pipelined mode, the average decoding time between two code words has been reduced from $n/p_s+2t+n/p_c$ pulses in the non-pipelined mode to $n/p_s$ pulses. It is important to note that the pipelining technique is particularly advantageous in the BCH decoder block when dealing with a large number of code words. This is because the necessary stages for error correction are performed in parallel on multiple code words, resulting in a decreased average error correction execution time by the BCH decoder block.

In the design of a pipeline, it is important to determine the duration of each cycle based on the worst-case scenario for each stage, even though some stages may be capable of completing their tasks more quickly. This is applicable to the BCH decoder circuit, where the pipeline technique ensures uninterrupted input processing and allows for simultaneous calculation of syndromes for new codewords while correcting errors and producing the output for previous codewords. On the other hand, the non-pipeline version of the BCH decoder circuit can only perform one task at a time. In general, employing the pipeline approach in scenarios with a large number of codewords significantly reduces the overall completion time, increases the throughput capacity, and enhances the efficiency of the BCH decoder circuit. Similarly, in flash memories, utilizing the pipeline technique enables the BCH decoder circuit to concurrently process multiple sectors, thereby speeding up the error correction process in flash memories.

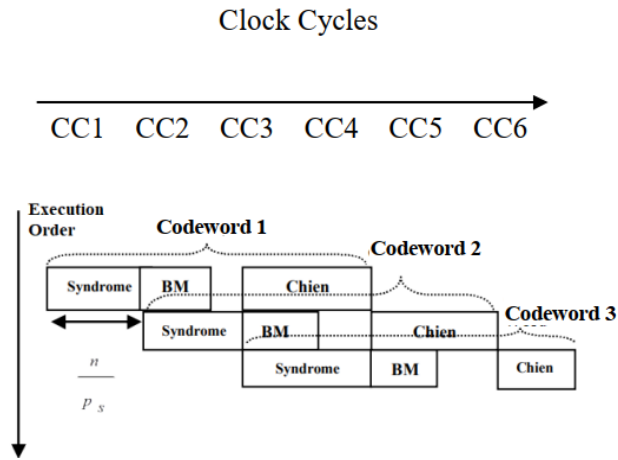

Figure 4: Pipeline Architecture of BCH Decoders

## 6. Simulation Results

All the proposed methods to decrease the time and hardware requirements of the BCH decoder block were tested and optimized using the 12.1 ISE Xilinx software. Initially, the design was written using a text editor that supports HDL and the VHDL hardware description language. After compiling the design, the desired VHDL program was synthesized using the FPGA programming tool, ISE Xilinx. This process generated circuit components like gates, flip-flops, adders, and a specific circuit netlist. The focus of this article is on implementing the error correction code (274,256) BCH for Flash NOR memories. It involves handling a 256-bit data page, which is composed of 16 words, each consisting of 16 bits.

Table 2: BCH Error Correction Code Parameters

| Parameters | Values |
|---|---|
| $m$ | $9[GF(2^9)]$ |
| $n$ | 274 |
| $k$ | 256 |
| $t$ | 3 |
| Redundancy bits | 18 |
| code rate (R=n/k) | 1.107 |

The chosen IC family for circuit synthesis in the ISE software is the FPGA 3Spartan series 1000S3XC. The table 3 displays the outcomes of synthesizing the encoder and decoder circuits for BCH.

Table 3: Parameters of BCH Error Correction Code

| Device Utilization Summary Selected Device: FPGA Sparatan3 XC3S1000 | | | | |
|---|---|---|---|---|
| Logic Utilization | Used Encoder | Encoder Utilization | Used **Decoder** | Decoder Utilization |
| Number of Slices | 11 out of 1920 | %1 | 41 out of 1920 | %2 |
| Number of Slice Flip Flops | 14 out of 3840 | %1 | 33 out of 3840 | %1 |
| Number of 4 input LUTs | 21 out of 3840 | %1 | 76 out of 3840 | %1 |
| Number of bonded IOBs | 14 out of 173 | %8 | 5 out of 173 | %2 |
| Number of GCLKs | 1 out of 8 | %12 | 1 out of 8 | %12 |

Table 4 presents the time required for encoding and decoding processes in each block of the BCH encoder and decoder. It also includes the operating frequency, critical path delay of sub-blocks in the BCH decoder, and the overall power consumption of the encoder and decoder circuits. The power consumption represents the sum of active and leakage powers, calculated using Compiler Power after synthesis. Based on Table 4, the minimum decoding time in the BCH decoder block is 2.5 nanoseconds. This corresponds to error-free received data, concluding the error detection process in the initial syndrome generation block. The average decoding time reflects when only one-bit error occurs in the received data, approximately 4.3 nanoseconds. The maximum decoding time is 7.9 nanoseconds, indicating the occurrence of the most errors in the received data, considering three bits in this article. The decoding process time varies depending on the number and location of errors in the stored data. The critical path delay in the BCH decoder block mainly results from finite field multipliers. Notably, sub-blocks such as search Chien and error locator polynomial generation contribute to increased delay. Employing the pipelining technique in the BCH decoder circuit enhances hardware resources and operational power, leading to a doubled acceleration in the decoding and error correction process between codewords. For the parameters mentioned in this article (n=274, p=4, t=3), the average decoding time between two codewords decreases from 143 clock pulses to 68 clock pulses.

Table 4: Executive specifications of BCH decoder and encoder circuit

| Parameters | Values | |
|---|---|---|
| Parallelization Factor (P) | 4 | |
| Encoding time | 4.3 *ns* | |
| Decoding time | 2/5 *ns* (min) | Reliable |
| | 4/ 3 *ns* (typ) | 1-Bit error |
| | 7/9 *ns* (max) | 3-Bit error |
| Working frequency | 200 MHz | |
| The combined power consumption of the BCH encoder and decoder circuit *(Active Power + Leakage Power)* | 7.32 μW | 1-Bit error |
| | 8.78 μW | 2-Bit error |
| | 11.7 μW | 3-Bit error |
| The delay of the critical path in each individual sub-block of the circuit responsible for BCH Decoder Circuit | Block Syndrome Production $T_{mult}+T_{add}+T_{ff} = 1/5$ *ns* | |
| | Debugger Polynomial Production Block $T_{mux}+T_{mult}+T_{add}+T_{ff} = 3/7$ *ns* | |
| | Chien Search Block $T_{mux}+T_{mult}+T_{add} = 4/2$ *ns* | |

The outcomes of implementing the XOR gate sharing multiplier to decrease the quantity of these gates in the parallel architecture of the Chien search block in the (274, 256) BCH decoder circuit are presented in Table 5. The findings demonstrate a noteworthy reduction in the number of XOR gates within the parallel architecture of the Chien search block after employing the sharing multiplier. Specifically, the overall reduction amounts to 23% upon sharing the XOR gates in each of the finite field multiplier units. It is worth noting that this reduction reaches 44% when the XOR gates are shared among a specific group of finite field multiplier units. Consequently, the sharing multiplier has the potential to significantly decrease the hardware complexity of the BCH decoder block.

Table 5: The outcomes derived from the XOR-sharing multiplier within the Chien search block

| Search Chien block architecture type | The number of XOR gates | Improvemen Rate |
|---|---|---|
| Serial Architecture | 102 | - |
| Parallel Architecture | 554 | - |
| An architecture that utilizes parallelism by incorporating XOR gates within each multiplier | 424 | % 23 |
| Parallel architecture which is based on sharing XOR gates between 4 groups of multipliers | 309 | % 44 |

The LFSR-based encoder block in this project requires approximately 4/3 nanoseconds to generate error correction bits. However, based on the timing specifications of a Flash NOR memory, as shown in Table 6, the maximum time required to write 32 bytes (256 bits) of data on a memory page is 128 microseconds. Therefore, implementing the BCH encoder block using an LFSR structure would be suitable for Flash NOR memories. In terms of accessing the stored data in the memory, even in the worst-case scenario with the highest number of errors in the received data, it would only require approximately 7/9 nanoseconds. This time is still lower than the maximum possible delay for sequential or random access in a Flash NOR memory, according to Table 6. Hence, based on the obtained results, the implemented BCH decoder block in this project, which utilizes a parallel structure with two sub-blocks (syndrome generation and Chien search) and the BM multiplier for error-locator polynomial generation, would be well-suited for the internal structure of the Flash NOR memory controller. This encoding and decoding process for error correction during data read and write operations will not cause any faults in the memory's functionality.

Table 6: Timing specifications of Flash NOR memory [52]

| Values | Timing specifications of Flash NOR memory |
|---|---|
| 128 μs/(32 Byte, 256 Bps) | Data Writing Time |
| 1s/ (128 KB) | Data Removing Time |
| 0.75 μs | Random Data Acquisition Time |
| 0.25 MB/s | Writing Bandwidth |

To assess the performance of the suggested decoder, it was compared to other methods in this article. The decoder was specifically designed to correct 3-bit errors and store 256 bits of data reliably in NOR Flash memory. The comparison, presented in Table 7, shows that the proposed pre-design method significantly improves the hardware complexity of the BCH decoder by reducing the number of XOR gates. The total number of XOR gates used in the BCH decoder circuit is 2400, which is half the number used in the other decoders mentioned in Table 7, demonstrating the superior hardware efficiency of the proposed decoder. In terms of decoding speed, the decoder introduced in this article operates faster than one of the compared decoders but slower than another. The relatively slower performance can be attributed to the use of the BM multiplier for error locator polynomial generation. While the BM multiplier has lower hardware complexity, its division operation, which is time-consuming in the finite field, contributes to the slower processing. In summary, the proposed BCH decoder in this article not only reduces decoding time and improves error correction but also achieves better hardware efficiency compared to other approaches.

Table 7: The comparative outcomes of the suggested BCH decoder and alternative approaches

| Parameter | Suggested BCH Decoder | BCH Decoder [12] | BCH Decoder [13] | BCH Decoder [26] |
|---|---|---|---|---|
| Implementation type | Synthesis FPGA Spartan 3 XC3S1000 | Post-Layout Technology 90 nm CMOS | Synthesis Technology 180 nm CMOS | Post-Layout Technology 90 nm CMOS |
| BCH $(n, k, t)$ | BCH (279 ،256 ،3) | BCH (279 ،256 ،3) | BCH (279 ،256 ،2) | BCH (255 ،239 ،2) |
| The number of XOR Gates | 2/4 K | 4/1 K | N/A | 4/2 K |
| Delay | 4/1 $ns$ (143 clock) | 2/5 $ns$ | 4/51 $ns$ | 6/8 $ns$ (272 clock) |

## 7. Conclusion

This article focuses on examining BCH codes and their utilization in encoding and decoding data in flash memories. The reliability of storage is a crucial challenge in the newer generation of flash memories with multi-level cell technology. Increasing the storage capacity by adding more levels per cell has resulted in a decrease in reliable data storage. Flash memories employ BCH error correction codes to ensure data integrity and detect/correct errors. These error correction codes, including various encoding and decoding multipliers, are extensively used in flash memory controllers.

Two fundamental aspects are present in the process of data decoding in these types of memories: enhancing computation speed and reducing the complexity of the hardware circuitry in the decoder. The data decoding process involves four steps: syndrome calculation, error locator polynomial calculation, the search Chien multiplier for error location detection, and error correction. This article introduces the implementation of a parallel architecture to minimize the delay caused by computations in each of the sub-blocks responsible for generating syndromes and performing the search Chien multiplier. Additionally, it suggests the utilization of an XOR gate sharing multiplier to eliminate redundant gates in the search Chien block. These strategies significantly contribute to improving the efficiency and hardware complexity of the BCH decoder circuit. Furthermore, employing a pipeline technique in the BCH decoder circuit enhances the utilization of hardware resources and operational power, resulting in a two-fold acceleration of the decoding and error correction process for each code word. The simulation results obtained using the VHDL hardware description language and the synthesis process with Xilinx ISE software confirm the anticipated enhancements achieved through these architectures and techniques when compared to other existing methods.

## References

1. K. Kim, "Future memory technology: challenges and opportunities," Proceedings of International Symposium on VLSI Technology, Systems and Applications, pp. 5-9, 2008.
2. K. Pangai, "90 nm Multi-level-cell flash memory technology," IEEE International Symposium on Semiconductor Manufacturing, pp. 197-199, 2010.
3. S. Lin and D. J. Costello, Error Control Coding: Fundamentals and Applications, Prentice-Hall Inc, 2004.
4. J. Cho and W. Sung, "Strength-reduced parallel chien search architecture for strong BCH codes," IEEE Transaction on Circuits and Systems, vol. 55, no. 5, pp. 427-431, 2008.
5. Y. M. Lin, C. H. Yang, C. H. Hsu, H. C. Chang and C. Y. Lee, "A MPCN-based parallel architecture in BCH decoders for NAND flash memory devices," IEEE Transaction on Circuits and Systems, vol. 58, no. 10, pp. 682-686, 2011.
6. H. Yoo, Y. Lee and I. C, Park, "7.3 Gb/s universal BCH encoder and decoder for SSD controllers," Design Automation Conference (ASP-DAC), pp. 37-38, 2014.
7. Y. Lee, H. Yoo, I. Yoo and I. C. Park, "High-throughput and low-complexity BCH decoding architecture for solidstate drives," IEEE Transactions on Very Large Scale Integration (VLSI) Systems, vol. 22, no. 5, pp. 1183–1187, 2014.
8. M. Yin, M. Wie and B. Yi, "Optimized multiplier for binary BCH codes," Circuits and Systems (ISCAS) IEEE Internationa Symposium, pp. 1552-1555, 2013.
9. D. V. Sarwate and R. S. Shanbhag, "High-speed architecture for BCH decoders," IEEE Transactions on Very Large Scale Integration (VLSI) Systems, vol. 9, no. 5, pp. 641–655, 2001.
10. H. Choi, W. Liu and W. Sung, "VLSI implementation of BCH error correction for multilevel cell NAND flash memory," IEEE Transactions on Very Large Scale Integration (VLSI) Systems, vol. 18, no. 5, pp. 843–847, 2010.
11. T. Chen, Y. Hsiao and Y. Hsing, "An adaptive-rate error correction scheme for NAND flash memory," 27th IEEE VLSI Test Symposium, pp. 53-58, 2009.
12. C. Chu, Y. Lin, C. Yang and H. Chang, "A fully parallel BCH codec with double error correcting capability for NOR flash applications," in IEEE International Conference of Acoustics, Speech, and Signal Processing, pp. 1605–1608, 2012.

13. X. Wang, P. Liyang, W. Dong and H. Chaohong, "A highspeed two cell BCH decoder for error correcting in mlc NOR flash memories," IEEE Transactions on Circuits and Systems, vol. 56, no. 11, pp. 865 – 869, 2009.
14. V. Mahadevaswamy, S. Sunitha and BN. Shobha, "Implementation of fault tolerant method using BCH code on FPGA," International Journal of Soft Computing and Engineering (IJSCE), vol. 2, no. 4, pp. 2231-2307, 2012.
15. Y. Jiang, A Practical Guide to Error-Control Coding Using MATLAB, Artech House, 2010.
16. F. Sun, S. Devarajan, K. Rose and T. Zhang, "Design of onchip error correction systems for multilevel NOR and NAND flash memories," IET Circuits, Devices and Systems, vol. 1, no. 3, pp. 241-249, 2007.
17. Y. Chen and K.K. Parhi, "Area efficient syndrome calculation for strong BCH decoding," Electronics Letters, vol. 47, no. 2, pp. 107-108, 2011.
18. W. Liu, J. Rho and W. Sung, "Low-power high-throughput BCH error correction VLSI design for multi-level cell NAND flash memories," IEEE Workshop on Signal Processing Systems Design and Implementation, pp. 303- 308, 2006.
19. L. Song, M. Yu and M. S. Shaffer, "10- and 40-Gb/s Forward error correction devices for optical communications," IEEE Journal of Solid-State Circuits, vol. 37, no. 11, pp. 1565-1573, 2002.
20. H. Lee, "High-speed VLSI architecture for parallel reedsolomon decoder," IEEE Transactions on Very Large Scale Integration (VLSI) Systems, vol. 11, no. 2, pp. 288–294, 2003.
21. M. Edoardo, VLSI Architecture for Computations in Galois Fields, Ph.D. Thesis, University of Linkoping, Sweden, WA, 2006.
22. C. Paar, "Optimized arithmetic for reed-solomon encoders," in Proceedings of IEEE International Symposium Information Theory, 250-250, 1997.
23. M. Potkonjak, M. Srivastava and A. Chandrakasan, "Multiple constant multiplications: efficient and versatile framework and multipliers for exploring common subexpression elimination," IEEE Transactions on Computer-Aided Design, vol. 15, pp. 151-165, 1996.
24. J. Hennessy and D. Patterson, Computer Architecture a Quantitative Approach, Morgan Kaufmann Publishers, 2007.
25. Micron, Technical Note: Parallel NOR Flash Embedded Memory.
26. Y. M. Lin, H. C. Chang and C. Y. Lee, "Improve high coderate soft BCH decoder architecure with one extra error compensation," IEEE Transactions on Very Large Scale Integration (VLSI) Systems, vol. 21, no. 11, pp. 2160–2164, 2013.